\documentclass[12pt]{article}
\usepackage{graphicx}
\usepackage{epsfig}%
\usepackage{graphicx}%
\usepackage{color}

\title{\centerline \bf Thermodynamics of the Lee-Wick partners: An
  alternative approach} \author{Kaushik Bhattacharya$^{\$}$,
  Suratna Das$^{\dagger}$
  \thanks{email:\,\,\,$^{\$}$kaushikb@iitk.ac.in,
    $^\dagger$suratna@tifr.res.in} \\ \normalsize $^{\$}$Department
  of Physics, Indian Institute of Technology, Kanpur, \\ \normalsize
  Kanpur 208016, India \\ $^\dagger$ Tata Institute of Fundamental
  Research\\ \normalsize Homi Bhabha Road, Colaba, Mumbai 400005,
  India\footnote{Major part of the work has been carried out while at
    Theoretical Physics Division, Physical Research Laboratory,
    Navrangpura, Ahmedabad 380009, India}}
\begin{document}
\maketitle
\begin{abstract}
It was pointed out some time ago that there can be two variations in
which the divergences of a quantum field theory can be tamed using the
ideas presented by Lee and Wick. In one variation the Lee-Wick
partners of the normal fields live in an indefinite metric Hilbert
space but have positive energy and in the other variation the Lee-Wick
partners can live in a normal Hilbert space but carry negative
energy. Quantum mechanically the two variations mainly differ in the
way the fields are quantized.  In this article the second variation of
Lee and Wick's idea is discussed. Using statistical mechanical methods
the energy density, pressure and entropy density of the negative
energy Lee-Wick fields have been calculated. The results exactly match
with the thermodynamic results of the conventional, positive energy
Lee-Wick fields. The result sheds some light on the second variation
of Lee-Wick's idea. The result seems to say that the thermodynamics of
the theories do not care about the way they are quantized.
\end{abstract}
\section{Introduction}
The Lee-Wick field theories \cite{Lee:1969fy,Lee} originated in an
attempt to address the problem related with the infinities in quantum
field theories.  Recently some authors have tried to implement
Lee-Wick's idea in a higher derivative version of a quantum field
theory \cite{Grinstein,Carone:2008bs}. All these theories assumed the
existence of some partners of the Standard model particles. The main
ideas of the Lee-Wick Standard Model in Ref.~\cite{Grinstein} have
been extended in Ref.~\cite{Carone1} where the authors use two
Lee-Wick partners for each standard model field: one with negative and
the other with positive norm. Later, this idea has been used in
Ref.~\cite{Carone2} to improve gauge coupling unification without
introducing additional fields in the higher-derivative theory. The
Higgs sector of the Lee-Wick Standard Model has also been constrained
in Ref.~\cite{Carone3}.

There has been at least one attempt \cite{Cai:2008qw} to use the
concepts of these Lee-Wick constructions in cosmology where the
authors were able to show the bouncing nature of the universe whose
energy is dominated by the energies of a scalar field and its Lee-Wick
partner. In Ref.~\cite{Fornal} the authors tried to formulate a
possible thermodynamic theory of particles which includes the Lee-Wick
partners using a method of statistical field theory previously
formulated by Dashen, Ma and Bernstein in Ref.~\cite{Dashen}.

The present article mainly focusses on a variant of the original
Lee-Wick idea,\footnote{Some work in this direction was started much
  before by Pauli in Ref.~\cite{Pauli}} which was concerned about the
taming of the divergences in a quantum field theory.  In 1984 Boulware
and Gross \cite{gross} tried to show that the original proposal of Lee
and Wick was related to a complex implementation of the Pauli-Villars
regularization scheme \cite{villars}. The complexity of the idea
arises from the fact that the Pauli-Villars regulator fields in the
Lee-Wick theories are not just ad hoc regulator fields, they also have
dynamics. To implement the Pauli-Villars idea Lee and Wick introduced
massive partner fields for all the normal fields in the theory. The
scheme becomes involved when one tries to quantize the partner
fields. It turns out that the partner fields can be quantized in two
ways. In one way the norm of the states of the partner fields on the
underlying Hilbert space remains definite and in the other case the
norm of the states of the partner fields on the Hilbert space becomes
indefinite. In the former case the energy of the Lee-Wick partner
fields turns out to be negative and in the later case the energy of
the Lee-Wick partner fields remain positive.

Both the options, as stated above, have their merits and demerits. In
the first option, where the partner field states have positive
definite norms but negative energy, the theory remains quantum
mechanically understandable but does not have a proper ground
state. There are run-away solutions. In the other option, where the
Lee-Wick partner field states live on an indefinite metric Hilbert
space but carry positive energy, there are zero-norm states which can
grow indefinitely. This option also gives rise to run-away
solutions. Historically Lee and Wick preferred to work with the theory
defined on an indefinite metric. The difficulties of the run-away
solutions were addressed by applying future boundary conditions which
again made the theory non-causal. 

The present article deals with the option which Lee and Wick
discarded, a quantum field theory of Lee-Wick fields whose states do
live on a definite metric Hilbert space but has negative energy. The
motivation for such an unconventional work comes from a very
interesting result related to the thermodynamics of the standard
Lee-Wick theory as given in Ref.~\cite{Fornal}. It turns out that the
thermodynamics of the indefinite metric, positive energy Lee-Wick
partners is exactly the same as definite metric but negative energy
Lee-Wick fields. This similarity of the thermodynamics of the two
different scenarios gives us a glimpse of the path which Lee and Wick
did not take historically. 

There are a plethora of problems related with the option which is
presented in this article, the most important of them being that the
theory is energetically unstable.  Presently we do not give all the
pathological properties of the alternative Lee-Wick prescription, nor
do we know the cures of all the formal (pathological) diseases of
the theory. The formal aspects of the negative energy Lee-Wick sector
remains mostly open for further investigation in the near future.
Inspite of all the conceptual difficulties related to run-away
solutions the result presented is too strong to be taken as a
coincidence.  Interestingly, the energy instability of the model plays
an important role in the thermodynamics of the unusual fields and
indirectly affects the results presented in this article which match
surprisingly with the thermodynamics of the positive energy Lee-Wick
fields.  Readers who are purely interested in the formal aspects of
the alternative Lee-Wick prescription can go through Ref.~\cite{gross}
for a more lucid and formal development of the basic ideas.

The present article is presented in the following manner. A brief
introduction on the unusual regulator fields and their properties is
presented in the next section.  Section \ref{distrb:s} discusses the
technique to find out the thermal distribution function of the
regulator fields. In section \ref{eps} the energy density, pressure
and entropy density of a gas comprising of elementary particles and
their unusual field partners are calculated using the thermal
distribution functions. The last section \ref{conc} summarizes the
important results obtained in this article.
\section{Canonical quantization of the positive energy and negative energy
Lee-Wick fields}
\label{gow}
In the paper written by Grinstein, O'Connell and Wise,
\cite{Grinstein} on Lee-Wick standard model, the authors proposed a
higher derivative field theory as the underlying theory of nature. The
quadratic kinetic terms of the normal field theories, both bosonic and
fermionic, are regained by introducing new degrees of freedom. It
turns out that the Lagrangians of the new fields have wrong signs.  In
their work Grinstein, O'Connell and Wise \cite{Grinstein} did not give
any prescription for the canonical quantization of the new fields. In
a later work by Fornal, Grinstein and Wise \cite{Fornal} the authors
derived the thermodynamics of the unusual new degrees of freedom using
methods of statistical field theory.  In this section we first
canonically quantize these new partner fields. This exercise will
immediately show that these fields do carry positive energy. In the
next step we give the prescription for obtaining negative energy
fields in the Lee-Wick paradigm.
\subsection{Indefinite metric positive energy case}
We do the quantization for the scalar field with the understanding
that the other bosons in the theory do follow the same quantization
rules. If the Lagrangian of the Lee-Wick partner of an usual scalar
field be represented as $\xi$, then according to \cite{Grinstein}, the
non-interactive part of its Lagrangian is given as
\begin{eqnarray}
{\mathcal L_\xi}= -\frac12 \partial_\mu\xi \partial^\mu\xi + 
\frac12 m_\xi^2\xi^2\,,
\label{tlag}
\end{eqnarray}
where $ m_\xi$ is the mass of the $\xi$ field. The field $\xi$ can
be expanded, in the Fourier space, in the same fashion as a standard scalar field:
\begin{eqnarray}
\xi(x) &=& \int \frac{d^3 p}{\sqrt{(2\pi)^3 2\epsilon({\bf p})}}
\left[a({\bf p})e^{-ip\cdot x} + \bar{a}({\bf p})e^{ip\cdot x}\right]\,,
\label{xiexp}
\end{eqnarray}
where $\epsilon({\bf p})=\sqrt{{\bf p}^2+ m_\xi^2}$ is the
dispersion relation of the Lee-Wick partner excitation. The partner
field can be quantized by the following condition:
\begin{eqnarray}
[\xi(t,{\bf
  x})\,,\,\pi_{\xi}(t,{\bf y})]=i\delta^3({\bf x-y})\,,
\end{eqnarray}
where $\pi_\xi\equiv \delta L_\xi/\delta \dot{\xi} = -\dot{\xi}$. The
quantization condition for the partner fields yields the following
unusual commutation relation
\begin{eqnarray} 
\left[a({\bf p})\,,\,\bar{a}({\bf q})\right]=-\delta^3({\bf p}-{\bf q})\,,
\label{nquanta}
\end{eqnarray}
while the other commutators involving $a({\bf p})$ and $\bar{a}({\bf
  q})$ are all zero. The equation above predicts that the field
excitations of the partner fields will have indefinite norm in the
Hilbert space. The unusual commutation relation between the creation
and annihilation operators of the $\xi$ field excitations is related to
the negative sign of the canonically conjugate momentum corresponding
to the $\xi$ field.  

If eigenstates of the number operator,
\begin{eqnarray}
N({\bf p})= -\bar{a}({\bf p})a({\bf p})\,, 
\label{dnop}
\end{eqnarray}
are defined by the following way
\begin{eqnarray}
N({\bf p})|n({\bf p})\rangle=n({\bf p})|n({\bf p})\rangle\,,
\label{nop}
\end{eqnarray}
where $n({\bf p})$ is a positive integer interpreted as the number of
particles with momentum ${\bf p}$ then the Hamiltonian of the field 
configuration is
\begin{eqnarray}
H=\int \epsilon({\bf p})N({\bf p})\,d^3 p\,, 
\label{hxi}
\end{eqnarray}
where we have dropped the zero-point contribution in the
Hamiltonian. The important point to note is the negative sign in the
definition of the the number operator. In this case if we stick to the
conventional definition of the number operator (the same operator
without the negative sign) then it can be shown that it will have
negative eigenvalues.  The quantity $ -\bar{a}({\bf p})a({\bf p})$ has
a positive spectrum. This unconventional behavior of the number
operator originates from the unconventional commutation relation of
the creation and annihilation operators of the field excitations as
given in Eq.~(\ref{nquanta}). The negative sign of the original
Lagrangian of the $\xi$ field is balanced by the negative sign of the
number operator and consequently the Hamiltonian of the field turns
out to be positive.

For the fermionic case one can take the Lagrangian of the Lee-Wick
partner field to be
\begin{eqnarray}
{\mathcal L}= -{\psi}^\dagger\gamma^0(i{\rlap /\partial}
-m_\psi)\psi\,,
\label{flag}
\end{eqnarray}
where $m_\psi$ is the mass of $\psi$ field excitations\footnote{The
  interested reader can consult 
  Refs.~\cite{sudarshan, Arons:1965wf} related to fermions leaving in an
  indefinite metric}.  The negative sign of the Lagrangian in
Eq.~(\ref{flag}) gives the unusual sign in conjugate momentum. The
fermionic field can be expanded in the Fourier basis as
\begin{eqnarray}
\psi(x) &=& \int \frac{d^3 p}{\sqrt{(2\pi)^3 2\epsilon({\bf p})}}
\sum_{s=1,2}\left[a_s({\bf p})u_s({\bf p})e^{-ip\cdot x} + 
\bar{b}_s ({\bf p})v_s({\bf p})e^{ip\cdot x}\right]\,,
\label{psiexp}
\end{eqnarray}
where $a_s({\bf p})$ and $b_s({\bf p})$ are the annihilation
operators of the fermionic and anti-fermionic excitations of the
$\psi$ field. Quantizing the fermion field $\psi$ in the conventional
sense
\begin{eqnarray}
\left\{\psi(t,{\bf x}), \pi_{\psi}(t,{\bf y})\right\}
=\delta^3({\bf x -y})\,,
\label{fermq}
\end{eqnarray}
one gets
\begin{eqnarray}
a_s^2({\bf p})=\bar{a}_s^2({\bf p})=0\,,\,\,\,\,
\left\{a_s({\bf p})\,,\,\bar{a}_{s'}({\bf k})\right\}=-\delta_{s,s'}
\delta^3\left({\bf p}-{\bf k}\right)\,.
\label{fquant}
\end{eqnarray}
In the fermionic case one can define the number operator for the
fermions and anti-fermions exactly in the same way as given in
Eq.~(\ref{nop}) and the form of the Hamiltonian will be similar to
that in Eq.~(\ref{hxi}). In this case also the Lee-Wick excitations
will have positive energy but indefinite norm. An excellent exposition
of the relationship between the unusual commutation relations of the
creation and annihilation operators (of the bosonic/fermionic fields)
and the indefinite metric it induces in the Hilbert space is presented
in the first two appendices of Ref.~\cite{Lee}.

In Ref.~\cite{Fornal} the authors tried to envisage the Lee-Wick
partners as intermediate resonances. According to them the normal
scalar particles scatter with each other as $\phi({\bf p}_1) +
\phi({\bf p}_2) \to \phi({\bf p'}_1) + \phi({\bf p'}_2)$ through two
Lee-Wick resonances. The scattering in reality happens like $\phi({\bf
  p}_1) + \phi({\bf p}_2) \to \xi(\tilde{{\bf p}}_1) + \xi(\tilde{{\bf
    p}}_2)$ and then $\xi(\tilde{{\bf p}}_1) + \xi(\tilde{{\bf p}}_2)
\to \phi({\bf p'}_1) + \phi({\bf p'}_2)$ where $\xi(\tilde{{\bf
    p}}_1)$ and $\xi(\tilde{{\bf p}}_2)$ stands for the Lee-Wick
fields. In this formalism the Lee-Wick partners are unstable
resonances, with a negative decay width, and their existence is
ephemeral. Writing the $S$-matrix as $S=1-i\mathcal{T}$ one can write
the $\mathcal{T}$ matrix amplitude for $\phi({\bf p}_1) + \phi({\bf
  p}_2) \to \xi(\tilde{{\bf p}}_1) + \xi(\tilde{{\bf p}}_2)$ as

\begin{eqnarray}
\langle \tilde{{\bf p}_1},\tilde{{\bf p}_2}|\mathcal{T}(E)|
{\bf p}_1,{\bf p}_2\rangle
=2\pi \delta(E_1+E_2-\epsilon_1 - \epsilon_2)
\delta^3({\bf p}_1+{\bf p}_2-\tilde{{\bf p}_1}-\tilde{{\bf p}_2})
\mathcal{M}(E)\,,
\end{eqnarray}
where the energy $E=E_1+E_2$ and momentum ${\bf p}={\bf p}_1+{\bf
  p}_2$ are such that $(p_1+p_2)^2= m^2_\xi$ which sets the threshold
value for the creation of the Lee-Wick resonances. Here $\epsilon_i$
stands for the energy of the Lee-Wick fields and $m_\xi$ is the mass
of the Lee-Wick excitations.  The important ingredient in
Ref.~\cite{Fornal} lies in the prescription used for writing
$\mathcal{M}(E)$:
\begin{eqnarray}
\mathcal{M}(E)=-\frac12 \frac{g^2}{E^2-{\bf p}^2-m^2_\xi+im_\xi \Gamma}\,,
\end{eqnarray}
where $g$ specifies the coupling of the normal fields with the
Lee-Wick fields. The interesting part of the above prescription lies
in the overall minus sign of $\mathcal{M}(E)$ and the negative sign
in front of the decay width $\Gamma$ in the denominator. The authors of
Ref.~\cite{Fornal} then utilize a conventional relation between the
scattering matrix elements and the grand partition function of a
thermodynamic system to predict the various thermodynamic
parameters. Thermodynamics of ephemeral resonance fields with negative
decay widths is highly non-trivial but in principle one can find it
out as done in Ref.~\cite{Fornal}.
\subsection{When the states have positive definite norm but negative energy}

In the present case we will try to find out the thermodynamics of the
negative energy Lee-Wick fields which live in a normal Hilbert space.
To do this one can start from the same Lagrangian for the real scalars
as used in Ref.~\cite{Grinstein} and as given in Eq.~(\ref{tlag}). The
field expansion of the $\xi$ field can be kept exactly the same as in
Eq.~(\ref{xiexp}).  To have a definite metric starting with a negative
Lagrangian one has to quantize the fields with the wrong sign as
\begin{eqnarray}
[\xi(t,{\bf
  x})\,,\,\pi_{\xi}(t,{\bf y})]=-i\delta^3({\bf x-y})\,.
\label{quantn}
\end{eqnarray}
This quantization condition yields the usual commutation relation
between the the creation and annihilation operators of the $\xi$ field
excitations. The $\xi$ field excitations will now lie in a normal
Hilbert space (i.e. the states having definite norm). The price one
has to pay for a cross over from the indefinite norm landscape to the
positive norm landscape is the ground state.  If we stick to the
definition of the number operator as defined in Eq.~(\ref{dnop}) then
$N({\bf p})$ will have negative eigenvalues.  Consequently the
Hamiltonian of the field configuration can still be written as
\begin{eqnarray}
H=\int \epsilon({\bf p})N({\bf p})\,d^3 p\,, 
\nonumber
\end{eqnarray}
but unlike the previous case the Hamiltonian will be having negative
eigenvalues. Although this Hamiltonian  will have zero as an
eigenvalue, but unlike the indefinite norm case, this eigenvalue is
not the least but the maximum of the eigenspectrum. Consequently the
definite norm Lee-Wick field excitations do not have a proper ground
state. If we interpret Eq.~(\ref{dnop}), for the indefinite metric
case, as the sum of positive energy excitations then  Eq.~(\ref{dnop})
has to be interpreted as a sum of positive energy de-excitations of the 
definite metric field. 

From a very similar analysis it can be shown that if one changes the
quantization condition for the fermions from that given in
Eq.~(\ref{fermq}) to
\begin{eqnarray}
\left\{\psi(t,{\bf x}), \pi_{\psi}(t,{\bf y})\right\}
=-i\delta^3({\bf x -y})\,,
\label{fermqn}
\end{eqnarray}
and defines the number operator for the particles as $N({\bf p})=
\bar{a}({\bf p})a({\bf p})$, which has positive eigenvalues, one
gets a Hamiltonian as
\begin{eqnarray}
H=-\int \epsilon({\bf p})N({\bf p})\,d^3 p\,. 
\end{eqnarray}
The negative sign in the Hamiltonian arises because the Lagrangian
of the fermionic field as given in Eq.~(\ref{flag}) carries a negative
sign.  Similarly the anti-fermions also carry negative energy. In this
article we are following the convention of Lee and Wick as given in
the first appendix in Ref.~\cite{Lee}. In their convention the
fermionic algebra can be set by a set of two anticommutation relations
(one for the positive definite norm, the other for the indefinite norm)
for the creation and annihilation operators of the fermions and
correspondingly there will be two expressions of the number
operator. According to Lee and Wick a fermionic system can be
quantized by the following set of rules:
\begin{eqnarray}
a_s^2({\bf p})=\bar{a}_s^2({\bf p})=0\,,\,\,\,\,
\left\{a_s({\bf p})\,,\,\bar{a}_{s'}({\bf k})\right\}=\pm\delta_{s,s'}
\delta^3\left({\bf p}-{\bf k}\right)\,,\,\,\,
N({\bf p})=\pm \bar{a}_{s}({\bf p}) \,a_s({\bf p})\,.
\label{convf}
\end{eqnarray}
When dealing with the indefinite metric theory we chose the minus
sign in the anticommutator and the minus sign in the number
operator. On the other hand when we are dealing with fermion fields
whose states live in a normal Hilbert space we choose the positive
sign of the anticommutator and the positive signed number operator.
\section{Thermal distribution functions of the regulator fields}
\label{distrb:s}
If some one assumes the existence of such exotic negative energy
Lee-Wick fields, may be during the earliest phases of the universe,
then one can calculate the thermodynamics of the negative energy
Lee-Wick fields.  The calculation of the thermodynamics of the
negative energy fields starts with the prediction of a thermal
distribution function of such fields. In this section the thermal
distribution function of both the bosonic and the fermionic Lee-Wick
partners are calculated.

For the bosonic excitations of the negative energy Lee-Wick fields we
know that the number operator and Hamiltonian has the form as given in
Eq.~(\ref{dnop}) and Eq.~(\ref{hxi}). In the present case the
Hamiltonian of the fields as given in Eq.~(\ref{hxi}) has negative
eigenvalues because the fields are assumed to be quantized with the
wrong sign as in Eq.~(\ref{quantn}).  Due to the presence of on-shell
excitations the thermal vacuum becomes $|\Omega\rangle\equiv |n({\bf
  p_1}),n({\bf p_2}),\cdots\rangle$ where $n({\bf p_1})$ is the number
of excitations carrying momentum ${\bf p_1}$. The action of the number
operator on such a vacuum is
\begin{eqnarray}
N({\bf p})|\Omega\rangle=n({\bf p})|\Omega\rangle\,.
\label{thermal-vacuum}
\end{eqnarray}
As the value of $n({\bf p})$ actually turns out to be negative for the
kind of Lee-Wick theory we are considering so in reality $|n({\bf
  p})|$ gives the number of particles with momentum ${\bf p}$ and
energy $\epsilon({\bf p})$ which are missing from the thermal
vacuum. In this analysis we will consider a non-interacting real
scalar field for which the chemical potential $\mu=0$.  In the
present case with a Hamiltonian of the form as given in
Eq.~(\ref{hxi}), where the number operator has negative eigenvalues,
the single particle partition function will be,
\begin{eqnarray}
z_{\rm LW}^B={\rm Tr}e^{-\beta H}=\sum_{|n({\bf p})|=0}^{\infty}e^{\beta \left|n({\bf p})\right|\epsilon(\bf p)},
\label{zlw}
\end{eqnarray}
where $\beta=\frac{1}{T}$. From the last equation it is seen that the
series representing the single particle partition function for the
Lee-Wick partner of a Standard model boson does not converge for
$\beta > 0$. Consequently we regularize the last expression by cutting
off the summation for a finite value of $n({\bf p})$ as:
\begin{eqnarray}
z_{\rm LW}^B=\sum_{|n({\bf p})|=0}^{M-1}
e^{\beta \left|n({\bf p})\right| \epsilon({\bf p})}=\frac{1 - e^{\beta 
\epsilon({\bf p}) M}}{1-e^{\beta \varepsilon({\bf p})}}\,,
\label{z2}
\end{eqnarray}
where $M$ is a dimensionless cut-off which can be made indefinitely
big at the end of the calculation.

Next we calculate the thermal distribution function of the field
excitations from the expression of the single cell partition function
of the field excitations as given in Eq.~(\ref{z2}).  In conventional
statistical mechanics we can find the single cell distribution
function via
\begin{eqnarray}
f({\bf p})=\frac{1}{\beta}\left(\frac{\partial \ln z}
{\partial \mu}\right)_{V,\beta}\,,
\end{eqnarray}
where $\mu$ is an auxiliary chemical potential whose exact nature is
not important for our purpose. In presence of an auxiliary chemical potential
the single particle partition function can be written as
\begin{eqnarray}
z_{\rm LW}^B=\sum_{|n({\bf p})|=0}^{M-1}e^{\beta \left|n({\bf p})\right|\{\epsilon({\bf p}) - \mu\}} =
\frac{1 - e^{\beta\left\{\epsilon({\bf p}) - \mu\right\}M}}
{1-e^{\beta \left\{\epsilon({\bf p}) - \mu\right\}}}\,.
\end{eqnarray}
Applying conventional methods, the distribution function can
also be written as
\begin{eqnarray}
f_{\rm B}({\bf p})=\frac{1}{\beta}\left(\frac{\partial \ln z^B_{\rm LW}}
{\partial \mu}\right)_{V,\beta}\,,
\label{LW_stat}
\end{eqnarray}
which comes out to be,
\begin{eqnarray}
f_{\rm B}({\bf p})=
-\frac{e^{\beta\left\{\epsilon({\bf p})-\mu\right\}}}
{1-e^{\beta\left\{\epsilon({\bf p})-\mu\right\}}} + 
M\frac{e^{\beta\left\{\epsilon({\bf p})-\mu\right\}M}}
{1-e^{\beta\left\{\epsilon({\bf p})-\mu\right\}M}} \,.
\end{eqnarray}
Now setting the auxiliary chemical potential to be zero we
get the distribution function of the fields as:
\begin{eqnarray}
f_{\rm B}({\bf p})= -\frac{1}{e^{-\beta \epsilon({\bf p})}-1}
+ \frac{M}{e^{-\beta \epsilon({\bf p}) M}-1}\,.
\label{lwpb}
\end{eqnarray}
\begin{figure}[h!]
\centering
\includegraphics[width=13cm,height=8cm]{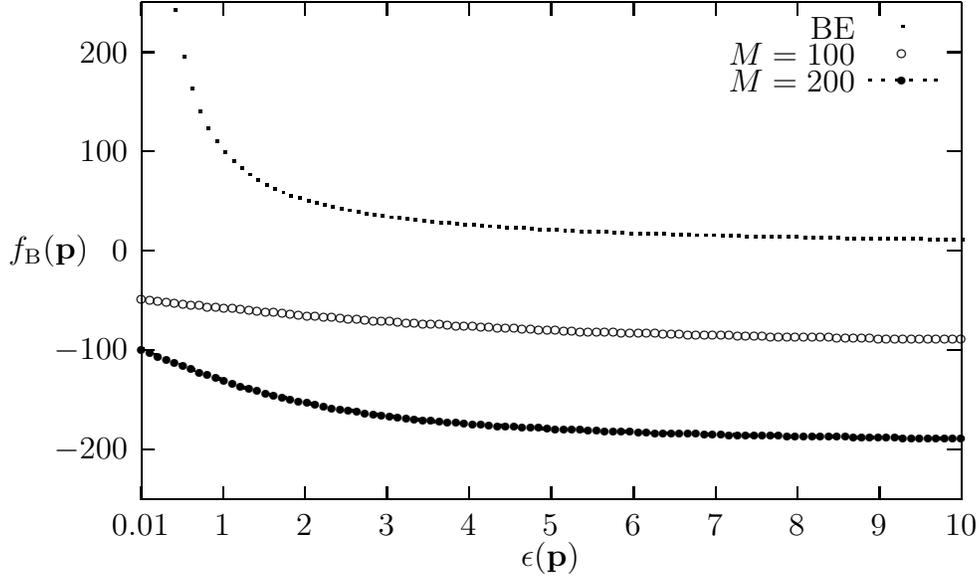}
\caption[]{The plot of the distribution function as given in
  Eq.~(\ref{lwpb}).  The topmost curve is for a normal Bose-Einstein
  distribution and the lower two curves correspond for $f_{\rm B}({\bf
    p})$ for the ultraviolet cutoff $M=100$ and $200$. The inverse
  temperature in all the cases is $.01{\rm GeV}^{-1}$ and the energy
  $\epsilon({\bf p})$ is in GeV.}
\label{lwb}
\end{figure}
This is the distribution function of the fields whose Lagrangian is
as given in Eq.~(\ref{tlag}). These fields are quantized via
Eq.~(\ref{quantn}). These are not the fields which appear in the
Standard model of particle physics. The distribution function as
plotted in Fig.~\ref{lwb} shows that the average excitation per energy
level is negative definite. Obviously these systems describe a
physical theory which is non-trivial and the negative sign of the
distribution is only meaningful when compared with the positive
definite distribution of the normal Standard model bosons. In general
these kind of distributions will produce negative energy density and
pressure but once these energy density and pressure is added with the
positive energy density and pressure of the Standard model bosons we
get a net positive energy density and pressure. The important point to
notice about the distributions is that there is no pile up of quantas
near $\epsilon({\bf p})=0$ as is in the case of the Bose-Einstein
distribution. The reason being that the spectra of the Lee-Wick
excitations with negative energy, as given in Eq.~(\ref{lwpb}), has
two infinite spikes as the energy tends to zero and they cancel each
other near the origin. 

If we take the Lagrangian of the fermionic fields as given in
Eq.~(\ref{flag}) and quantize them via Eq.~(\ref{fermqn}) then the
anticommutators of the creation and the annihilation operators, which
define the fermionic excitations of the $\psi$ field discussed in the
last section, is given as
\begin{eqnarray}
a_s^2({\bf p})=\bar{a}_s^2({\bf p})=0\,,\,\,\,\,
\left\{a_s({\bf p})\,,\,\bar{a}_{s'}({\bf k})\right\}=\delta_{s,s'}
\delta^3\left({\bf p}-{\bf k}\right)\,,
\end{eqnarray}
where $s$, $s'$ may be some internal quantum numbers. We can
proceed in a similar way as done before and calculate the thermal
distribution of these excitations. If we use the conventional 
number operator $N({\bf p})\equiv \bar{a_s}({\bf p})a_s({\bf p})$,
which has positive eigenvalues as 0 and 1, then the
Hamiltonian of a single oscillator is
\begin{eqnarray}
H({\bf p})=-\frac12\epsilon({\bf p})\left[\bar{a}_s({\bf p}){a}_s({\bf p})
-{a}_s({\bf p})\bar{a}_s({\bf p})\right]=-\epsilon({\bf p})\left[N({\bf p})-
\frac12\,\delta^3({\bf 0})\right]\,,
\label{phamf}
\end{eqnarray}
where $\frac12\,\epsilon({\bf p})\,\delta^3({\bf 0}) $ is the zero
point energy.  The negative sign of the Hamiltonian is due to the
negative sign of the Lagrangian of the $\psi$ field.  A similar
analysis can be done for the anti-fermions also. If we use this
oscillator Hamiltonian to calculate the single particle partition
function there will be no problem related to the convergence of the
series. The single particle partition function for the anticommuting
fields turns out to be
\begin{eqnarray}
z_{\rm LW}^F=\sum_{n({\bf p})=0}^{1}e^{\beta n({\bf p})\{\epsilon({\bf p}) - \mu\}}=
1+e^{\beta \left\{\epsilon({\bf p}) - \mu\right\}}\,,
\end{eqnarray}
where $\mu$ is an auxiliary chemical potential. Now applying the
formula in Eq.~(\ref{LW_stat}) and setting $\mu=0$ at the end we get
the distribution function for the definite metric Lee-Wick partners of
the Standard model fermions as:
\begin{eqnarray}
f_{\rm F}({\bf p})=-\frac{1}{e^{-\beta\epsilon({\bf p})}+1}\,.
\label{lwpf}
\end{eqnarray}
\begin{figure}[h!]
\centering
\includegraphics[width=13cm,height=8cm]{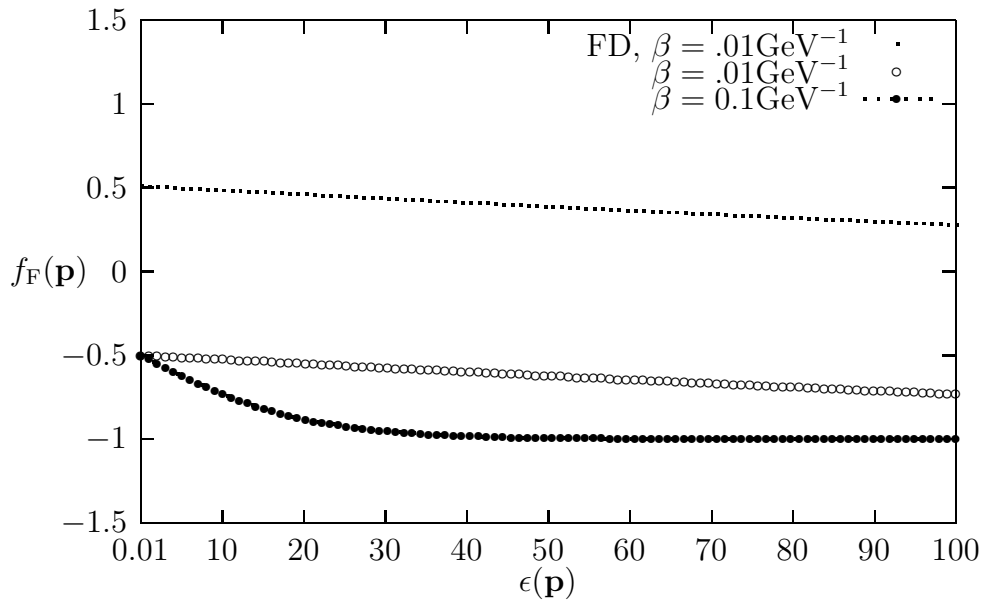}
\caption[]{The plot of the distribution function as given in
  Eq.~(\ref{lwpf}).  The topmost curve is for a normal Fermi-Dirac
  distribution at $\beta=.01{\rm GeV}^{-1}$ and the lower two curves
  correspond for $f_{\rm F}({\bf p})$ for $\beta$ values $.01{\rm
    GeV}^{-1}$ and $.1{\rm GeV}^{-1}$  and the energy
  $\epsilon{({\bf p})}$ is in GeV.}
\label{lwf}
\end{figure}
Unlike the previous case, in the present scenario the distribution
function has no dependence on the dimensionless regulator $M$. 

The negative sign of the distribution function signifies that the
present field configurations arises due to a de-excitation or loss of
positive energy particles. The vacuum defined is not stable and there
exists much less energetic states than the vacuum itself. These kind
of fields are unstable. The maximum energy of the field configurations
is zero.
\section{Energy density, pressure and entropy density from the distribution 
function.}
\label{eps}
\subsection{The bosonic case}
To calculate the relevant thermodynamic quantities for the bosonic
field from a statistical mechanical point of view we will employ
Eq.~(\ref{lwpb}). The energy density can be calculated using the
following known equation
\begin{eqnarray}
\rho = \frac{g}{\left(2\pi\right)^3}\int\epsilon({\bf p})f_{\rm B}
(\epsilon)d^3p
=\frac{g}{2\pi^2}\int_0^\infty \epsilon({\bf p})f_{\rm B}(\epsilon)
|{\bf p}|^2d{|\bf p}|\,,
\end{eqnarray}
where $g$ stands for any intrinsic degree of freedom of the particle.
For a relativistic excitation $\epsilon^2={\bf p}^2+m^2$ where $m$ is
the mass of the bosonic excitations. Changing the integration variable
from $|{\bf p}|$ to $\epsilon$ one gets
\begin{eqnarray}
\rho=-\frac{g}{2\pi^2}\int_0^\infty\left(\epsilon^3-\frac{m^2}{2}
\epsilon\right) \frac{d\epsilon}{e^{-\beta \epsilon}-1} + 
\frac{Mg}{2\pi^2}\int_0^\infty\left(\epsilon^3-\frac{m^2}{2}
\epsilon\right) \frac{d\varepsilon}{e^{-\beta\epsilon M}-1}\,.
\end{eqnarray}
In the above integral it is assumed that $|{\bf p}|\gg m$ and to have
a closed integral the lower limit of the integral is assumed to be
zero. In the extreme relativistic limit the system temperature $T \gg m$.
Both of the integrals can only be done when $\beta < 0$, and in that
case the result of the last integral is
\begin{eqnarray}
\rho=-\frac{g}{2\pi^2}\left(\frac{\pi^4T^4}{15}-
\frac{m^2\pi^2T^2}{12}\right) + \frac{g}{2\pi^2}
\left(\frac{\pi^4T^4}{15M^3}-
\frac{m^2\pi^2T^2}{12M}\right)\,.
\end{eqnarray}
Analytically continuing the above result for $\beta>0$ and taking $M
\to \infty$ we see that for normal temperatures the energy density for
extreme relativistic excitations of the bosonic fields is of the
following form
\begin{eqnarray}
\rho= -g\left(\frac{\pi^2T^4}{30}-\frac{m^2T^2}{24}\right)\,.
\label{endenb}
\end{eqnarray}
As expected, the energy density turns out to be negative for the
excitations in this case. The pressure of the bosonic field
excitations can be found out from
\begin{eqnarray}
p &=&\frac{g}{\left(2\pi\right)^3}\int\frac{|{\bf p}|^2}{3\epsilon}
f_{\rm B}(\epsilon) d^3p=\frac{g}{2\pi^2}\int_0^\infty
\frac{|{\bf p}|^4}{3\epsilon}
f_{\rm B}(\varepsilon)d|{\bf p}|\,.
\end{eqnarray}
Following similar steps as in the case of the energy density, it is
seen that the pressure of extremely relativistic excitations of the
bosonic fields turns out to be
\begin{eqnarray}
p= -g\left(\frac{\pi^2T^4}{90}-\frac{m^2T^2}{24}\right)\,.
\label{pb}
\end{eqnarray}
The entropy density of the bosonic field is simply given by
\begin{eqnarray}
s=\frac{\rho + p}{T} =-g\left(\frac{2\pi^2T^3}{45} - \frac{m^2T}{12}\right)\,.
\end{eqnarray}
These values of the energy density, pressure and entropy density
exactly match the corresponding values calculated for the
conventional, positive energy Lee-Wick partners. In \cite{Fornal} the
authors were trying to formulate thermodynamics for a
higher-derivative theory. The higher derivative theory was converted
into standard theory (theory up to a second derivative) with the
introduction of Lee-Wick partners whose states have indefinite
norms. The authors in the previous work did not quantize the system
explicitly but were working with the form of the propagators of the
Lee-Wick partners.

If we assume that in the early universe for each bosonic degrees of
freedom in the Standard model there exist a corresponding Lee-Wick bosonic
degree of freedom whose field configuration has negative energy then
the net energy density, pressure and entropy density of the early
universe turns out to be
\begin{eqnarray}
\rho_{B} = \rho_{\rm SM} + \rho = \frac{g m^2T^2}{24}\,,\,\,\,\,
p_{B} = p_{\rm SM} + p = \frac{g m^2T^2}{24}\,,\,\,\,\,
s_{B} = s_{\rm SM} + s=\frac{gm^2T}{12}\,,
\label{tboson}
\end{eqnarray}
 which are all positive as expected. Here the energy density, pressure
 and entropy density for the Standard model bosonic particles are
 $\rho_{\rm SM}= g\frac{\pi^2T^4}{30}$, $p_{\rm
   SM}=g\frac{\pi^2T^4}{90}$ and $s_{\rm SM}=g\frac{2\pi^2T^3}{45}$
 respectively \cite{kolb}.
\subsection{The fermionic case}
In this subsection we apply Eq.~(\ref{lwpf}) to find the energy
density, pressure and entropy density of the fermionic excitations. In
this case the distribution function do not have any dependence on the
regulator $M$. For relativistic excitations the integrals which give
the energy density and pressure for the fermionic case are exactly
similar with the bosonic case except that now we have to use the
distribution for the fermions. The integrals can be easily done,
granted $\beta < 0$, but the results can be analytically continued for
positive temperatures. The results in this case are listed below. The
energy density, pressure and entropy density of the Lee-Wick partners
are as follows:
\begin{eqnarray}
\rho&=&-g\left(\frac{7\pi^2T^4}{240}-\frac{m^2T^2}{48}\right)\,,
\label{rhofi}\\
p &=&-g\left(\frac{7\pi^2T^4}{720}-\frac{m^2T^2}{48}\right)\,,
\label{pf}\\
s&=& -g\left(\frac{7\pi^2 T^3}{180} - \frac{m^2}{24}\right)\,.
\label{sfi}
\end{eqnarray}
The energy density and pressure quoted above are equivalent to the
energy density and pressure for the positive energy Lee-Wick partners
as calculated in \cite{Fornal} for the special case
of $g=2$. If we assume that to each unusual fermionic degree of
freedom there corresponds one standard fermionic degree from the
Standard model, then the total fermionic contribution is
\begin{eqnarray}
\rho_F = \rho_{\rm SM} + \rho = \frac{g m^2T^2}{48}\,,\,\,\,
p_F = p_{\rm SM} + p = \frac{g m^2T^2}{48}\,,\,\,\,
s_F = s_{\rm SM} + s = \frac{g m^2T}{24}\,
\label{sf}
\end{eqnarray}
which are all positive. Here the energy density, pressure density
and entropy density for the Standard model fermionic particles are $\rho_{\rm
  SM}=g\frac{7\pi^2T^4}{240}$, $p_{\rm SM}=g\frac{7\pi^2T^4}{720}$ and
$s_{\rm SM}=g\frac{7\pi^2T^3}{180}$ respectively \cite{kolb}.

It is worth pointing out here that there is some confusion regarding
higher derivative theories of fermions. The confusion is regarding the
number of Lee-Wick partners (one left-handed and the other
right-handed) of the chiral fermions. The authors of
Ref.~\cite{Fornal} claim that there will be two positive energy
Lee-Wick partners of a chiral fermion which are interrelated. Where as
in Ref.~\cite{Wise} the author claims that the two positive energy
Lee-Wick partners of the chiral fermion may not be interdependent. In
that case the Lee-Wick degrees of freedom exceeds the one of its
Standard model partner yielding negative energy, pressure and entropy
density. This issue is yet to be resolved.
\section{Discussion and conclusion}
\label{conc}
Initially it was pointed out that Lee-Wick's idea of implementing the
Pauli-Villars regularization scheme can be implemented in two ways.
This idea was presented by Boulware and Gross \cite{gross} way back in
1984.  In one way the regulator fields live in a indefinite metric
space but carry positive energy and in the other way the regulator
fields live in a definite metric space but carry negative energy. Lee
and Wick took the first option and tried to redress the issue of
indefinite norm in such a way that unitarity is preserved in the
theory. The second option remained uncultivated. In this article we
explored the second option with limited means. No cures for the energy
instability of these kind of theories are known to the present
authors. The results presented in the article show the dubious nature
of the energy instability of the fields, but instead of making the
theory meaningless the same instabilities conspire to produce a result
which matches with the thermodynamics of the Lee-Wick partners living
in the indefinite metric space.

In this article we have studied a system of bosonic and fermionic
fields, whose Lagrangians have the wrong sign and, which are quantized
with the wrong sign of the commutators and the anticommutators.  These
fields are Lee-Wick partners who live in a normal Hilbert space but
have negative energy excitations. The negative energy of the field
configuration is not due to any particular form of the potential but
solely an outcome of the negative sign of the Lagrangian and the
modified quantization process. The vacuum of the theory is not the
state with the lowest energy, it is rather the state with the maximal
energy making the field configuration unstable. The bosonic and
fermionic degrees of freedom do still follow commutation and
anticommutation relations and specifically the fermionic fields still
follow the Pauli exclusion principle. In this article the emphasize
had been on the calculation of energy density, pressure and entropy
density of the unusual field configurations.

To calculate the above mentioned thermodynamic quantities one requires
to have a statistical mechanics of the field excitations. One
encounters the difficulty of a diverging sum when calculating the
single particle partition function of the bosonic fields. Keeping to
conventional ways, where the temperature of the system is positive
definite, the partition function can only be summed when one uses an
ultraviolet cutoff. The distribution function calculated from the
partition function turns out to be negative definite, which is a
nontrivial result. The negative nature of the distribution function
implies that there must be an average loss of particles in any
energy level. 

The energy density, pressure calculated from the distribution
functions of the unusual fields discussed in this article match
exactly with the results calculated by Fornal, Grinstein and Wise in
\cite{Fornal}.  The derivation of the new distribution functions and
the connection between the regulator field thermodynamics as presented
in this article and the thermodynamics of the Lee-Wick partners, as
presented in Ref.~\cite{Fornal}, is one of the main motivations for
this work.  The main emphasize of the present article has not been to
recalculate the results obtained in Ref.~\cite{Fornal} as the theory
presented in this article is not the same as that of
Ref.~\cite{Fornal}. The two theories are quantized differently.  The
similarity of the thermodynamic results of the two variants of the
Lee-Wick model implies that the different kind of instabilities
plaguing the theories are related in presence of a thermal bath.  From
the main analysis of this article it can be inferred that the
conventional Lee-Wick prescription is equivalent to its variant in the
thermodynamic sector.  The analysis of the thermodynamics of the
positive definite metric Lee-Wick partners gives a clear idea about
the origin of the negative energy density and pressure of these field
configurations. The theory presented in the article is amenable to the
standard techniques of finite temperature field theory.  One can
utilize the results of finite-temperature quantum field theory to
calculate various thermal effects, like the thermal mass of the
Standard model particles in presence of the thermalized Lee-Wick
partners, in the early universe.  The thermal distribution functions
of the Lee-Wick partners calculated in the present article can be used
to write the propagators of the unusual fields in the real-time
formalism \cite{Nieves:1990ne}.

The fact that the thermodynamics of the indefinite metric, positive
energy Lee-Wick fields and the thermodynamics of the definite metric
but negative energy Lee-Wick partners turns out to be the same remains
an interesting result which invites further work in these fields in
the future.
\vskip .2cm
\noindent 
{\bf Acknowledgement}: We want to thank  Subhendra Mohanty
for pointing out some subtle ideas involved in this article.

\end{document}